\documentclass[10pt]{article}
\usepackage{algpseudocode}
\usepackage{amsmath}
\usepackage{algorithm}
\usepackage{graphicx}
\usepackage{cite}
\usepackage{rotating}
\usepackage{color}
\usepackage{subcaption}
\usepackage{booktabs}
\usepackage{enumerate}
\usepackage{tabularx}
\usepackage{algorithm}
\usepackage{verbatim}
\usepackage{pgf}
\usepackage{url}

\usepackage[labelfont=bf,labelsep=period,justification=raggedright]{caption}
\usepackage{lineno}

\makeatletter
\renewcommand{\@biblabel}[1]{\quad#1.}
\makeatother

\date{}

\begin{document}

\begin{flushleft}
{\Large
\textbf{Monte Carlo-based Noise Compensation in Coil Intensity Corrected Endorectal MRI}
}
\\
$Dorothy~Lui^{1}$,
$Amen~Modhafar^{1}$,
$Masoom~A.~Haider^{2}$,
$Alexander~Wong^{1,\ast}$
\\
\bf{1} Department of Systems Design Engineering, University of Waterloo, Waterloo, Canada
\\
\bf{2} Department of Medical Imaging, Sunnybrook Health Sciences Centre, Toronto, Canada
\\
$\ast$ E-mail: a28wong@uwaterloo.ca
\end{flushleft}

\section*{Abstract}

\textbf{Background}: Prostate cancer is one of the most common forms of cancer found in males making early diagnosis important. Magnetic resonance imaging (MRI) has been useful in visualizing and localizing tumor candidates and with the use of endorectal coils (ERC), the signal-to-noise ratio (SNR) can be improved. The coils introduce intensity inhomogeneities and the surface coil intensity correction built into MRI scanners is used to reduce these inhomogeneities. However, the correction typically performed at the MRI scanner level leads to noise amplification and noise level variations.
\textbf{Methods}: In this study, we introduce a new Monte Carlo-based noise compensation approach for coil intensity corrected endorectal MRI which allows for effective noise compensation and preservation of details within the prostate.  The approach accounts for the ERC SNR profile via a spatially-adaptive noise model for correcting non-stationary noise variations.  Such a method is useful particularly for improving the image quality of coil intensity corrected endorectal MRI data performed at the MRI scanner level  and when the original raw data is not available.
\textbf{Results}:  SNR and contrast-to-noise ratio (CNR) analysis in patient experiments demonstrate an average improvement of 11.7 dB and 11.2 dB respectively over uncorrected endorectal MRI, and provides strong performance when compared to existing approaches.
\textbf{Conclusions}: A new noise compensation method was developed for the purpose of improving the quality of coil intensity corrected endorectal MRI data performed at the MRI scanner level.  We illustrate that promising noise compensation performance can be achieved for the proposed approach, which is particularly important for processing coil intensity corrected endorectal MRI data performed at the MRI scanner level and when the original raw data is not available.\\~\\
\section{Background}
\label{sec:intro}

Prostate cancer (PCa) is one of the most commonly diagnosed cancers among North American men, encompassing an estimated $14\%$ and $24\%$ of all new cancer cases in the United States and Canada respectively. In 2014, an estimated $233,000$ American and $23,600$ Canadian men are expected to be diagnosed with PCa and of those cases, $29,480$ and $4,000$ are expected to result in death~\cite{US_PCaStat,CAD_PCaStat}. Prostate specific antigen (PSA) blood assay and digital rectal exams are exams used for screening PCa. High PSA levels indicate high PCa risk. The use of PSA is controversial and often inadequate as it over-detects clinically insignificant prostate cancer, resulting in a high degree of over-treatment. Treatment of prostate cancer with radiation or surgery carries significant risk of life altering side effects such as sexual dysfunction, urinary and rectal incontinence and thus should not be undertaken unless necessary~\cite{PSA,EDPC}. After a positive screening, the next step is systematic transrectal ultrasound (TRUS) guided biopsy which involves systematic regional sampling of the prostate with typically $8$ or more samples being taken. This is invasive and uncomfortable and suffers from sampling error as the tumors are not easily visible with TRUS. As such, it is important to consider detection alternatives. Magnetic Resonance Imaging (MRI) has been shown to be a viable alternative as it can visualize the cancer and has a good negative predictive value for significant cancer, helping avoid unnecessary biopsy and reduction of sampling error.

MRI has become a commonly used diagnostic imaging tool for detecting PCa due to its improved contrast between cancer and background healthy tissue in a tomographic view. Better signal-to-noise ratio (SNR) can be achieved using a localized surface receiver coil placed directly over the body region of interest (ROI) to increase the magnetic sensitivity. Placed on the skin surface, these surface coils are relatively far from the centrally located prostate (i.e. $> 10$ cm). Alternatively, endorectal coils (ERCs) placed in the rectum are within a few millimeters of the prostate gland. With both surface and ERCs, the signal decreases farther away from the coil and consequently introduces intensity inhomogeneities.  ERCs have recently been shown to offer a diagnostic advantage~\cite{Turkbey2013} in the detection of prostate cancer compared to surface coils at 3 T. As such, there remains a strong interest in utilizing ERC despite the discomfort associated with insertion of the endorectal balloon. For lower field systems operating at 1.5 T, an ERC is helpful in achieving performance similar to 3 T MRI with pelvic phased-array coils (PAC)~\cite{Sosna2004,Beyersdorff2005}. The results demonstrated no significant visualization difference between the two approaches, although according to Beyersdorff~et al.~\cite{Beyersdorff2005} ERCs exhibited improved SNR. Thus, the use of ERCs remains a particular interest at 1.5 T as well. Conversely, the ERC's inhomogeneous sensitivity results in high intensities at the prostate's peripheral zone nearest the coil and decreases in intensity near the upper region of the central gland, making visualization, delineation and diagnosis difficult~\cite{Lui2014}.

\begin{figure}
\centering
\includegraphics[width = 0.5\linewidth]{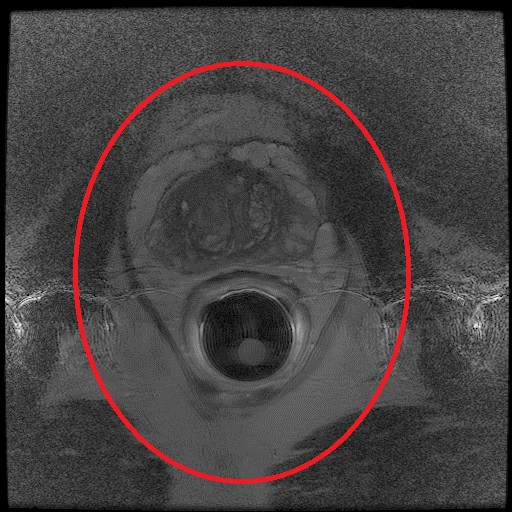}
\caption{An example of amplified noise in low SNR regions (ie. regions outside the red ellipse) following a pre-calibrated intensity bias correction approach. Exam was performed on a 1.5 T system using a Hologic endorectal receiver coil.}
\label{fig:noiseamp}
\end{figure}

Due to physiological limitations, ERCs are designed to be small, which causes inhomogeneous signal distribution. To counterbalance this, MRI scanners are equipped with coil intensity correction techniques that improve images through a post-reconstruction or a pre-calibration correction technique.  Built-in MRI pre-calibration correction approaches are often preferred for MRI acquired using ERCs when compared to post-reconstruction techniques as they provide more accurate bias field estimates which leads to more reliable corrected MRI images.  A pre-calibration correction approach proposed by Liney et al.~\cite{Liney1998} uses a series of proton-density (PD) weighted images acquired prior to the acquisition to generate the bias field estimate to be used for correction during the actual acquisition. This approach has been realized in commercial systems such as Phased array UnifoRmity Enhancement (PURE - General Electric (GE)), Prescan Normalize (Siemens), CLEAR (Philips) and NATURAL (Hitachi).  One of the consequences of using such intensity correction approaches is it creates a spatial dependence on background noise (which is uniformly distributed~\cite{SCICb,Liney1998} prior to correction). This results in increasing noise levels as we move away from the coil in the corrected images, which is particularly visible in regions distant from the coil~\cite{SCICb}. An example is shown in Fig.~\ref{fig:noiseamp} where the regions outside the red ellipse indicate low SNR regions where noise has been intensified as a result of pre-calibrated intensity bias correction. As such, a post-processing approach to address this noise amplification due to pre-calibration correction of coil intensity for endorectal MRI would be very beneficial, given its widespread use.  This is particularly useful for retrospective studies where the original raw data is not available and only the coil intensity corrected data is accessible.

MRI noise is an issue under active research~\cite{Wong2011,Shafiee}. It amounts to difficult analysis and hinders post-processing approaches such as segmentation and registration~\cite{Manjon2009, Golshan2013, phase1,phase2,phase3,orchard}. Raw MRI data is complex (both real and imaginary components) and represented in the frequency domain with additive Gaussian noise. Transforming this complex data to the spatial domain renders the magnitude data to be Rician distributed~\cite{Foi2011,Golshan2013,Gudbjartsson1995}. The data distribution is also dependent upon the SNR, where low SNR regions (mainly described by noise only) can be modeled as Rayleigh distributed~\cite{Gudbjartsson1995, Sijbers2004} and high SNR regions as Gaussian distributed~\cite{Gudbjartsson1995,Golshan2013,Wiest2008}. Moreover, the signal-dependent nature of noise in the intensity corrected images introduces challenges to noise compensation.

Taking the characteristic distributions of MRI data into consideration, noise can be compensated. Numerous approaches have been proposed using MRI magnitude data to compensate for noise, using a variety of methods including total variation~\cite{Keeling2011, Varghees2012, Martin2013}, analyzing multiple scales using wavelet denoising~\cite{Pizurica2003,Anand2010,Wang2012}, via non-local means~\cite{Coupe2011,Wiest2008,Manjon2009,Manjon2010,nlm} and linear minimum mean-square estimators (LMMSE)~\cite{Golshan2013,Aja2008}. These approaches combine a mixture of techniques to handle the particular nature of MRI noise: spatial-adaptation to the noise variance~\cite{Varghees2012, Wong2011, Coupe2011,Maximov2012}, Rician distribution~\cite{Varghees2012, Wang2012, Martin2013, Coupe2011, Awate2007} and accounting for signal-dependent bias when using a Gaussian assumption~\cite{Anand2010, Wang2012, Wong2011, Coupe2011}.

In this study, a new approach called Adaptive Coil Enhancement Reconstruction (ACER) is introduced that is suitable for coil intensity corrected endorectal MR images. ACER reconstructs noise-compensated endorectal MR magnitude images using a stochastic Bayesian estimation framework. A spatially-adaptive Monte Carlo sampling approach is introduced to estimate the posterior distribution using a Rician model. The Monte Carlo posterior estimation is modified to model the Rician-nature of MRI magnitude data. Moreover, the SNR profile of the specific ERC used is incorporated into the posterior estimation by integrating a learned parametric non-stationary Rician model. The model is learned using maximum likelihood estimation based on the data and specifications of the ERC. The posterior estimate is then used to form a noise-suppressed reconstruction using Bayesian least-squares estimation. Given the pressures of acquiring MRI data more quickly, the proposed approach offers an alternative to obtain increased SNR by post-processing retrospective coil intensity corrected data for improved visualization.

\section{Methods}
\label{method}

In this section, the problem is formulated and the process of how the noise-compensated image is reconstructed is discussed.

\subsection{Problem Formulation}
The acquired MRI magnitude image, $V$, can be expressed as the following relationship~\cite{Manjon2012}:

\begin{equation}\label{Eq1}
V(s) = G(s) + N(s)
\end{equation}

\noindent where $s$ is the pixel location, $G$ is the noise-compensated reconstruction and $N$ is the non-stationary noise. Knowing the noise process $N$, Eq.~\ref{Eq1} can be reformulated as an inverse problem where the noise-compensated reconstruction $G$ can be found. Bayesian least-squares estimation~\cite{Fieguth2010,Wong2011b,Milanfar2013} is used to estimate $G$ that minimizes the expected squared estimation error. This formulation is shown below:

{\small\begin{equation}\label{Eq2a}
\begin{aligned}
\hat G(s) &= \operatorname*{arg\,min}_{\hat G(s)} {E\left((G(s)-\hat G(s))^2|V(s)\right)}\\
& = \operatorname*{arg\,min}_{\hat G(s)} \left({\int(G(s)-{\hat G(s)})^2p(G(s)|V(s))dG(s)}\right)
\end{aligned}
\end{equation}}

\noindent Taking the derivative of Eq.~\ref{Eq2a}:

{\small\begin{equation}\label{Eq2b}
\begin{aligned}
&\frac{\partial}{\partial \hat{G}(s)}\int(G(s)-\hat{G}(s))^2 p(G(s)|V(s))dG(s)=\\
&\int \{-2(G(s)-\hat{G}(s))p(G(s)|V(s))dG(s)\}
\end{aligned}
\end{equation}}

\noindent Then setting the derivative in Eq.~\ref{Eq2b} to zero:
{\small\begin{equation}\label{Eq2c}
\begin{aligned}
\int G(s)p(G(s)|V(s))dG(s) &= \int \hat{G}(s)p(G(s)|V(s))dG(s)\\
&= \hat{G}(s)\int p(G(s)|V(s))dG(s)\\
&= \hat{G}(s)
\end{aligned}
\end{equation}}

\noindent Simplifying to:
\begin{equation}\label{Eq2d}
\hat{G}(s) = \underbrace{\int G(s)p(G(s)|V(s))dG(s)}_{E(G(s)|V(s))}
\end{equation}

\noindent In Eq.~\ref{Eq2d}, $G(s)$ can be estimated using the conditional mean of $G(s)$ on $V(s)$, $E(G(s)|V(s))$, or the mean of the posterior distribution, $p(G(s)|V(s))$. An estimate of the posterior distribution, $p(G(s)|V(s))$, can be calculated using a spatially-adaptive importance-weighted Monte-Carlo sampling approach. The approach is adapted to account for the non-stationary Rician characteristics of MRI magnitude data. This is explained in more detail in the next section.

\subsection{Spatially-Adaptive Rician Distributed Monte Carlo Posterior Estimation}\label{sec:mcmc}

\noindent MRI magnitude data is Rician distributed, following:

{\small\begin{equation}\label{Eq3a}
f(x|\nu,\Phi) = \frac{x}{\Phi^2}\exp\bigg(\frac{-(x^2 + \nu^2)}{2\Phi^2}\bigg)I_0\bigg(\frac{x\nu}{\Phi^2}\bigg), \quad x > 0; \nu, \Phi \geq 0,
\end{equation}}

\noindent where $\Phi$ and $\nu$ are parameters that control the distribution's scale and skew and $I_0$ is the modified Bessel function of the first kind with order zero. As a result of coil intensity correction, the data's Rician distribution becomes spatially-dependent and results in the following distribution, where $x > 0; \nu, \Phi \geq 0$:

{\small
\begin{equation}\label{Eq3b}
f(x|\nu(s),\Phi(s)) = \frac{x}{\Phi(s)^2}\exp\bigg(\frac{-(x^2 + \nu(s)^2)}{2\Phi(s)^2}\bigg)I_0\bigg(\frac{x\nu(s)}{\Phi(s)^2}\bigg)
\end{equation}}

This distribution can be accounted for in estimating the posterior distribution via an importance-weighted Monte Carlo sampling approach~\cite{Chen1994}. The approach forms $\Omega$, a set of samples and importance weights selected from a search space, $\eta$. Pixels, $s_k$, are selected in a region around a pixel of interest, $s_0$, and from these samples, a subset are collected randomly using an instrumental distribution, $Q(s_k|s_0)$, such as a uniform distribution. For each randomly drawn pixel, $s_k$, an acceptance probability, $\alpha(s_k|s_0)$~(Eq.~\ref{Eq3c}), is calculated which indicates the probability that the neighbourhood of $s_k$ is similar to the neighborhood of $s_0$:

{\small\begin{equation}\label{Eq3c}
\alpha(s_k|s_0) =  \frac {\prod_{j}\nolimits \frac{x(j)}{\hat{\Phi}(s_0)^2} \exp\big(\frac{-(x(j)^2+{\nu(j)}^2)}{2\hat{\Phi}(s_0)^2}\big)I_0\big(\frac{x(j){\nu(j)}}{\hat{\Phi}(s_0)^2}\big)}{\prod_{j}\nolimits \lambda}
\end{equation}}

\noindent where $x(j) = h_k[j]$ and $\nu(j) = h_0[j]$. The terms $h_k[j]$ and $h_0[j]$ denote the $j^{th}$ pixels in the neighbourhoods around $s_k$ and $s_0$. The variable $\lambda$ normalizes $\alpha(s_k|s_0)$ so that in the case the neighbours of $s_k$ are duplicates of $s_0$, $\alpha(s_k|s_0)=1$. The variables $\hat{\Phi}(s_0)$ is the estimated scale, for the pixel of interest, $s_0$ (its estimation is explained in more detail in the following section). This acceptance probability is used to determine if the sample $s_k$ is a realization of the posterior $p(G(s)|V(s))$ and should be accepted into the set $\Omega$. The acceptance probability reformulates the Rician-distributed statistics to handle the non-stationarity of the coil-intensity corrected MRI data when deciding whether a pixel is accepted or rejected. To use the acceptance probability, a random value $u$ is first generated from a uniform distribution. Then, the pixel $s_k$ is accepted into the set $\Omega$ if $u \le \alpha(s_k|s_0)$, otherwise it is rejected. The process of selection and acceptance is continued until $N$ samples are accepted into $\Omega$. The posterior distribution estimate can then be calculated using a weighted-histogram~\cite{Chen1994}:

{\small\begin{equation}\label{Eq3e}
\hat{p}(G(s)|V(s)) = \frac{\sum_{j \in \Omega}\alpha(s_j|s_0)\delta(G(s) - V(s_j))}{Z}
\end{equation}}

\noindent where $\delta()$ is the Dirac delta function and $Z$ is a normalization term to enforce $\int \hat{p}(G_j|V_j) = 1$. The posterior distribution can then be used to calculate the noise-compensated reconstruction $\hat{G}(s)$ using Eq.~\ref{Eq2d}.

\subsection{Non-Stationary Unified ERC Parametric Model}\label{sec:var}
To estimate the posterior distribution $p(G(s)|V(s))$ in a spatially-adaptive manner, the scale parameter of each pixel of interest, $\hat{\Phi}(s_0)$, is estimated using maximum likelihood estimation,

\begin{equation}\label{Eq4}
\hat{\theta}_{ML} = \arg \max_{\theta} f(x|\theta)
\end{equation}

\noindent where x are the observed intensities in $V(s)$ and $\theta$ are the parameters to be estimated: in this case, the scale parameter, $\hat{\Phi}(s_0)$. To refine the scale estimation, an existing SNR profile, defined as  $\gamma(\theta)$, which is characteristic to a given ERC, is fitted. Given an ERC, an SNR profile can be mapped to characterize the change in SNR as a function of distance from the ERC surface. Literature has shown that the ERC SNR profile differs from a rigid and inflatable coil, however both coils share a common trend where there is an SNR gain nearest the coil surface which diminishes with distance~\cite{Venugopal2010, Noworolski2008, Arteaga2012}. Considering the SNR depth profile from posterior to anterior of a rigid coil, a sharp increase in SNR of 3 to 5 times the normal SNR is demonstrated at the ERC surface. This increase is followed by a decrease through the peripheral zone and central gland. Despite the quick decline in SNR, the peripheral zone still experiences a gain in SNR of 1.5 to 3 times. The continual decrease then finds the central gland with only a fraction of the SNR~\cite{Noworolski2008, Venugopal2010, Arteaga2012}. An inflatable coil has demonstrated a weaker response with less SNR increase near the coil. In addition to the variation between SNR profiles for inflatable and rigid ERC, ERC brands have their own characteristic profiles which can be determined by measuring phantoms.  Two SNR profiles were modeled in this study using the findings from Venugopal et al.~\cite{Venugopal2010} for two ERCs: a Hologic rigid ERC and a Medrad inflatable ERC.  The inflatable and rigid ERC SNR profiles demonstrate a 1 and 5-fold improvement in SNR at the ERC surface respectively with an exponential drop leading to a final abrupt drop.  The full algorithm, Adaptive Coil Enhancement Reconstruction (ACER), is summarized in Algorithm~1.

\begin{algorithm}\label{algo:ACER}
\caption{ACER Algorithm Summary:}
\begin{algorithmic}
\State{1. Perform model fitting to estimate the local scale map, $\hat{\Phi}(s)$, using the ERC's SNR profile}
\State{2. Using the instrumental distribution, $Q(s_k|s_0)$, select a subset of pixels randomly from the neighbourhood of the pixel of interest $s_0$ in $V$}
\State{3. Calculate the acceptance probability, $\alpha(s_k|s_0)$, for each $s_k$ in the of subset selected pixels in step 2.}
\State{4. Select a random value, $u$, from a uniform distribution}
\If{$u <= \alpha(s_k|s_0)$}
	\State{The pixel, $s_k$, is considered a realization of $\hat{p}(G(s)|V(s))$ and is accepted into the set $\Omega$}
\Else
	\State{The pixel, $s_k$, is not a realization of $\hat{p}(G(s)|V(s))$ and is discarded}
\EndIf	
\State{5. Calculate the posterior distribution as a weighted histogram using $\alpha(s_k|s_0)$ for all $s_k$ in $\Omega$ (Eq.~\ref{Eq3e})}
\State{6. Use posterior distribution to calculate $\hat{G}(s_0)$ (Eq.~\ref{Eq2d})}
\State{7. Repeat steps 2 - 6 for each pixel in $V$}
\end{algorithmic}
\end{algorithm}

\subsection{Experiments}

To interpret the performance of the proposed approach, clinical patient data and phantom data were used. Clinical patient and phantom endorectal T2 (spin-spin relaxation time) and axial diffusion-weighted MRI (DWI) corrected with the pre-calibration coil intensity correction approach by GE called Phased array UnifoRmity Enhancement (PURE) was collected at Sunnybrook Health Sciences Center. Two types of coils were used to acquire the data: an inflatable Medrad eCoil ERC and a rigid Hologic ERC. The data was collected using a GE Discovery MR750 3 T MRI for phantom data (inflatable coil only) and a GE Signa HDxt 1.5 T MRI for patient data (collected with both rigid and inflatable ERCs).

The proposed approach was compared against three other MRI denoising approaches: 1.) an optimized variance-stabilizing transformation for Rician distributions (ROVST)~\cite{Foi2011}, 2.) noise removal by a multi-resolution adaptive non-local means approach (ANLM)~\cite{Coupe2011} and 3.) a linear minimum mean squared error estimator (LMMSE)~\cite{Aja2008}. The ROVST, LMMSE and ANLM codes used for comparison were provided by their respective authors. All approaches were implemented using MATLAB and the parameters were selected to provide a reasonable balance between prostate detail and noise compensation in the background. The experimental setup for the phantom and patient experiments are described in more detail in the following sections.

\subsubsection{Phantom Experimental Setup}	\label{sec:phant_exptSetup}
For phantom experiments, a multi-modality prostate training phantom from Computerized Imaging Reference Systems Inc (CIRCS Model 053) was used. The phantom is contained within a $12 \times 7.0 \times 9.5$ cm clear container made of acrylic. The container has two openings for the probe (front - $3.2$ cm diameter and rear - $2.6$ cm diameter). Located inside the container is a prostate replica composed of high scattering Blue Zerdine ($5.0 \times 4.5 \times 4.0$ cm) that is placed in a water-like background gel with little backscatter attenuation ($\leq 0.07$ dB/cm-MHz). Within the prostate itself, there are three $0.5 - 1.0$ cm lesions placed hypoechoic to the prostate. The urethra and rectal wall are made of low scattering Zerdine with diameter of $0.70$ cm and dimensions $6.0 \times 11 \times 0.5$ cm respectively.

\begin{figure}[ht!]
\centering
\includegraphics[width = 0.5\linewidth]{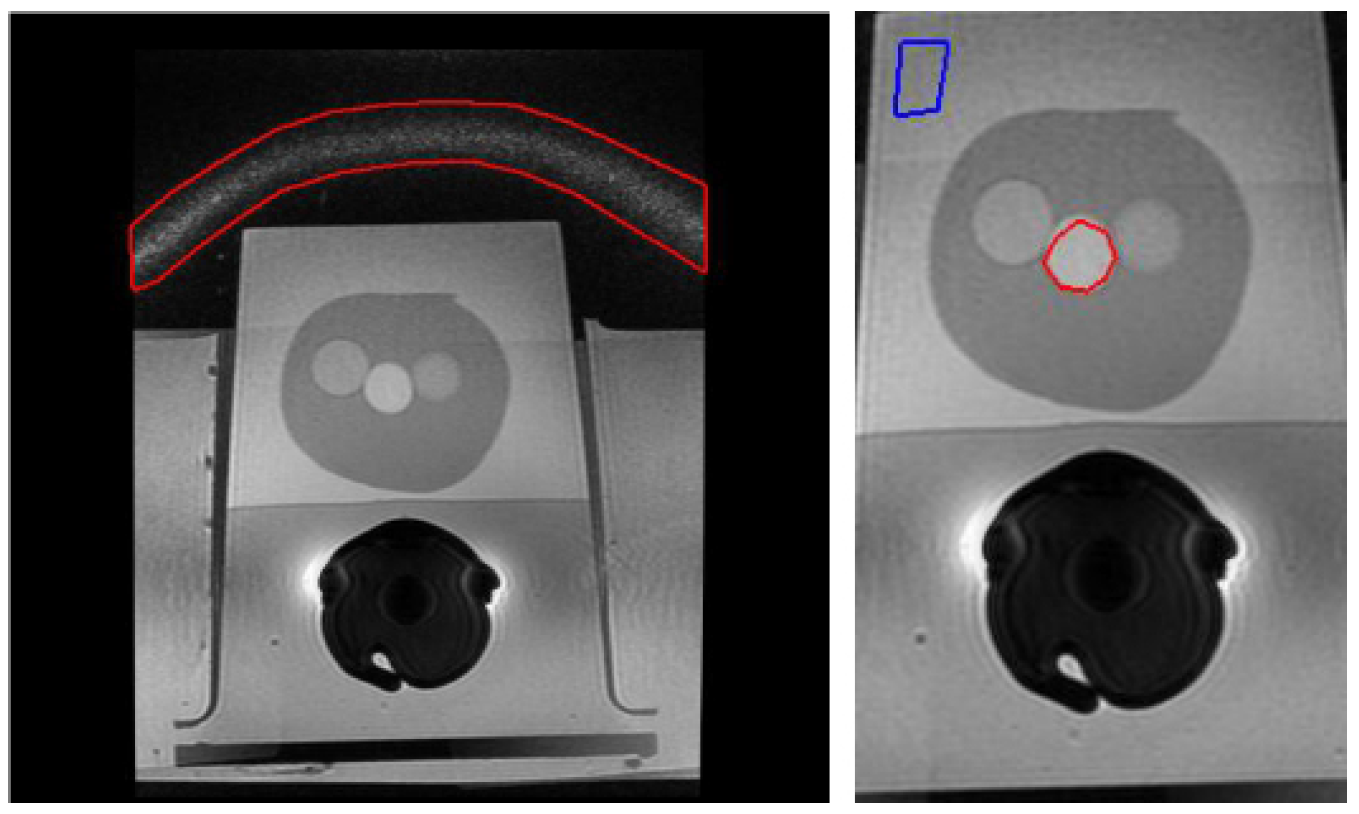}
\caption{Left: Uncorrected uncropped T2 slice. A noise band (red) is present due to PURE correction which amplifies the noise around the cushion used to stabilize the phantom during imaging. Right: Corresponding slice cropped for processing to include only the ROI with selected regions (blue and red) for SNR and CNR calculation on phantom DWI and T2.}
\label{fig:noise_phant}
\end{figure}

This phantom was then placed in a tub of water to increase signal amplification and placed between cushions to elevate and stabilize the phantom during acquisition and to improve the realism of the phantom. The phantom was then imaged with the inflatable Medrad Prostate eCoil MR endorectal coil using T2 MRI and DWI. Both T2 and DWI MRI were acquired with the built-in pre-calibration correction approach PURE using one excitation with a 3 T GE Discovery MR750. The three phantom data sets were acquired using 1) DWI $b = 0$ s/$\text{mm}^2$, 2) DWI $b = 1000$ s/$\text{mm}^2$ and 3) T2 and the central slice selected for experimentation. As a result of PURE correction, the cushion in these slices are emphasized by a noise band shown in Fig.~\ref{fig:noise_phant} for T2. To focus on the phantom itself, these slices were cropped.

The display field of view (DFOV) is $16 \times 16$ cm with a pixel spacing of $0.3$ mm between rows and columns for DWI acquisitions and $0.6$ mm between rows and columns for T2 acquisitions. Both DWI and T2 had common slice thicknesses of $3$ mm. The echo time for T2 was $107$ ms while the echo time for DWI was $72$ ms. The repetition time for T2 was $3,200$ ms and $10,000$ ms for DWI. Central slices from each modality were then considered for SNR and CNR.

\subsubsection{Patient Experimental Setup}\label{sec:pat_expt}
The second experiment evaluates the image reconstruction performance of the various tested approaches on endorectal T2 axial MRI with PURE within a clinical scenario. The data was collected and then selected for this study retrospectively using a GE Discovery 1.5 T Signa HDxt MRI scanner, a Medrad eCoil inflatable ERC or a Hologic rigid ERC. Institutional research ethics board approval and patient informed consent for this study was obtained. For the purpose of evaluating imaging reconstruction performance, fourteen patient cases were used in this study. Eleven patients were imaged using an inflatable Medrad coil and the central slices were selected for analysis. Three patients were imaged using a rigid Hologic coil and three slices were selected from each volume and considered as a separate case. The patients ranged in age from $54 - 79$ years with a median age of $72$ years. The data was collected using $0.5$ excitations (NEX) with echo times ranging from $100 - 107$ ms (median echo time of $104$ ms) and repetition times of $3,400$ ms. Each slice has a DFOV of $16 \times 16$ cm with a pixel spacing of $0.3$ mm between rows and columns and a slice thickness of $3$ mm. The central slices from each patient case were assessed using SNR, CNR and edge preservation and 3 cases were selected to be qualitatively assessed via a subjective scoring method.

\subsection{Results and Discussion}
\label{results}

Following the experimental setup, a number of quantitative and qualitative analysis methods were executed to evaluate the performance of the proposed approach against the state-of-the-art techniques.

\subsubsection{Phantom Experiment}\label{sec:expt_phant}
For the phantom experiments, the noise suppression approaches were compared using signal-to-noise ratio (SNR), contrast-to-noise ratio (CNR) and visual analysis. P-values were also calculated to determine the statistical significance of the SNR and CNR results. The null hypothesis used was that a given correction approach had no improvement for a subjective metric as compared to the uncorrected image. P-values were calculated for a two-tailed normal distribution with a statistical significance level of $5\%$.

\begin{figure*}[ht!]
	\centering
	\includegraphics[width=1\linewidth]{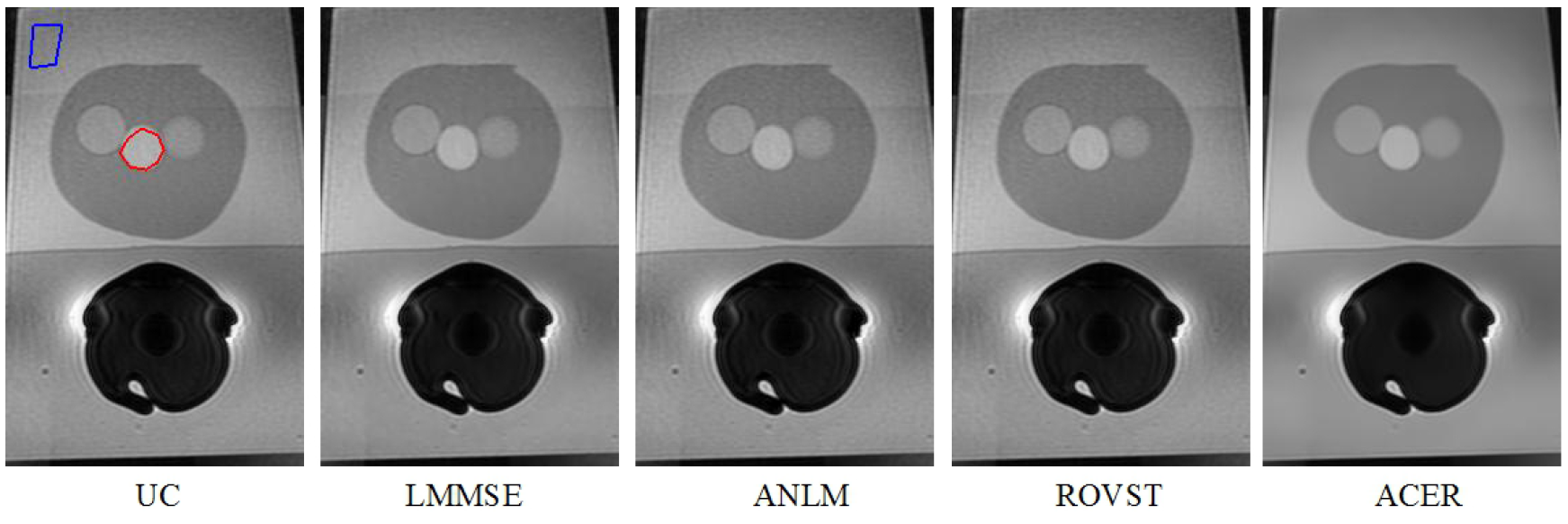}
	\caption{Noise suppressed T2 phantom experiment results: A background region (blue) and a prostate region (red) are shown where the SNR and CNR were calculated in the uncorrected (UC) slice. ACER maintains a good balance between noise compensation in smooth regions while retaining edges.}
	\label{fig:phant}
\end{figure*}

Due to the known homogeneity of the phantom, for quantitative analysis, SNR and CNR were calculated for two regions: one region on the phantom farthest away from the coil and a second region on the prostate itself. These regions are shown in Fig.~\ref{fig:phant} in the uncorrected image in blue and red respectively. SNR and CNR (in decibels) were calculated as follows:

\begin{equation}\label{Eq11}
SNR = 20\log\frac{\bar{x}}{\sigma}, \qquad CNR = 20\log\frac{|\bar{x_A}-\bar{x_B}|}{\sigma}
\end{equation}
%

In the SNR equation, the parameter, $\bar{x}$, defines the mean value of the region and $\sigma$ signifies the standard deviation of the region. In CNR, $\bar{x}_A$ and $\bar{x}_B$, denote the mean values of the selected background and prostate regions respectively and $\sigma$ is the standard deviation of the background region which is more indicative of the noise process.

The final SNR and CNR results are shown in Table~\ref{tab:snr_phant} and Table~\ref{tab:cnr_phant} with visual results for the T2 phantom case in Fig.~\ref{fig:phant}. All approaches demonstrated improvement upon the uncorrected (UC) slice with the proposed approach, ACER, having the highest average SNR in the selected background and prostate regions. The uncorrected (UC) slice refers to the slice with no application of any algorithm. ROVST and LMMSE proved to have the next best SNRs in the two regions however, considering the visual results, noise was under or overcompensated with deterioration of structure. In the case of DWI at $b = 1000~\text{s/mm}^2$, where noise was more prominent and contrast was already low, ROVST had greater SNR metrics over ACER however, at the cost of structure preservation. Finally, ANLM exhibited the least SNR improvement in both selected regions, indicating an inaccurate noise estimate.

\begin{table*}[htbp]
  \centering
  \caption{Phantom SNR analysis of a selected background and prostate region (in dB with highest measures in bold). ACER proved to have the greatest SNR improvement in the background  and prostate regions. ANLM showed an inaccurate noise variance estimate which led to less significant SNR improvement.}
    \begin{tabular}{c|ccccc|ccccc}
    \toprule
          & \multicolumn{5}{c|}{\textbf{Background SNR}} & \multicolumn{5}{c}{\textbf{Prostate SNR}} \\
    \midrule
    \textbf{Case} & \textbf{ACER} & \textbf{ROVST} & \textbf{LMMSE} & \textbf{ANLM} & \textbf{UC} & \textbf{ACER} & \textbf{ROVST} & \textbf{LMMSE} & \textbf{ANLM} & \textbf{UC} \\
    \midrule
    $DWI_{b=0}$ & \textbf{33.2} & 32.0  & 31.6  & 30.9  & 30.6  & \textbf{27.0} & 26.8  & 26.7  & 26.2  & 26.1 \\
    $DWI_{b=1000}$ & 27.5  & \textbf{27.6} & 26.4  & 26.0  & 25.9  & 27.3  & \textbf{27.5} & 26.9  & 25.9  & 25.7 \\
    T2    & \textbf{29.2} & 27.0  & 27.6  & 27.0  & 26.9  & \textbf{27.2} & 26.7  & 26.9  & 26.7  & 26.7 \\
	\midrule
    \textbf{Avg.} & \textbf{30.0} & 28.9  & 28.5  & 27.9  & 27.8  & \textbf{27.2} & 27.0  & 26.8  & 26.3  & 26.2 \\
    \bottomrule
    \end{tabular}%
  \label{tab:snr_phant}%
\end{table*}%

\begin{table}[htbp]
  \centering
  \caption{Phantom CNR analysis based on the selected background and prostate regions (in dB with highest measures in bold). ACER demonstrated the greatest improvement in CNR illustrating its capacity to augment the detail within the prostate.}
    \begin{tabular}{c|ccccc}
    \toprule
    \textbf{Case} & \textbf{ACER} & \textbf{ROVST} & \textbf{LMMSE} & \textbf{ANLM} & \textbf{UC} \\
    \midrule
    $DWI_{b=0}$ & \textbf{27.1} & 25.9  & 25.4  & 24.7  & 24.5 \\
    $DWI_{b=1000}$ & \textbf{20.9} & 21.0  & 19.7  & 19.4  & 19.4 \\
    T2    & \textbf{19.7} & 17.6  & 18.1  & 17.5  & 17.5 \\
    \midrule
    \textbf{Avg.} & \textbf{22.6} & 21.5  & 21.1  & 20.5  & 20.4 \\
    \bottomrule
   \end{tabular}%
  \label{tab:cnr_phant}%
\end{table}%

CNR analysis (Table~\ref{tab:cnr_phant}) showed that ACER had the highest average CNR. ROVST had the second highest average CNR and ANLM with the least improvement. These results indicate ACER's ability to increase the contrast between the background and prostate regions, thereby improving the visibility of detail within the prostate.

P-values (Table~\ref{tab:pval_phant}) were also calculated for the SNR and CNR results to verify the statistical significance of the differences between the average SNR and CNR compared to the uncorrected slices. ACER and LMMSE proved their greater average background SNRs over the uncorrected slices were statistically significant with p-value scores of less than $0.05$. ACER additionally demonstrated statistical significance for CNR. The other approaches instead had p-values greater than $0.05$ indicating the change over the uncorrected slices was not representative of any notable change. For the prostate SNR, all approaches demonstrated statistically insignificant results with p-values greater than $0.05$.

\begin{table}[htbp]
  \centering
  \caption{The p-values for the metrics measured for the phantom experiments. Values below $0.05$ are shown bolded which indicate the average score across the cases has statistical significance. ACER and LMMSE approaches are the only approaches to have statistically significant results over the uncorrected slices.}
    \begin{tabular}{c|cccc}
    \toprule
    \textbf{Metric} & \multicolumn{1}{c}{\textbf{ACER}} & \multicolumn{1}{c}{\textbf{ROVST}} & \multicolumn{1}{c}{\textbf{LMMSE}} & \multicolumn{1}{c}{\textbf{ANLM}} \\
    \midrule
    \textbf{Background SNR} & \textbf{0.02}  & 0.16  & \textbf{0.04}  & 0.27 \\
    \textbf{Prostate SNR} & 0.10  & 0.25  & 0.14  & 0.18 \\
    \textbf{CNR} & \textbf{0.02}  & 0.16  & 0.05  & 0.24 \\
    \bottomrule
    \end{tabular}%
  \label{tab:pval_phant}%
\end{table}%

The noise suppressed T2 phantom slices for each approach are shown in Fig.~\ref{fig:phant}. The proposed method demonstrates the best noise compensation while enhancing the detail contrast within the prostate. LMMSE and ROVST also compensate for noise however at the cost of visible structure and edge blurring.

\subsubsection{Patient Experiment}\label{sec:expt_real}
The noise suppression approaches were then compared using patient data by analyzing SNR, CNR (Eq.~\ref{Eq11}), edge preservation (Eq.~\ref{Eq9}) and subjective scores. P-value analysis was also included to determine the statistical significance of the results.

For the SNR and CNR assessment, a high noise, structure-free region in the background was selected similar to the phantom experiments. A second homogeneous region with higher intensity was then selected for CNR calculation. The results are shown in Table~\ref{tab:snr_cnr_real} where all approaches improved upon the background SNR of the uncorrected slice. In the case of background SNR, ACER had the highest average SNR with ROVST in second. The visual results for ROVST and LMMSE demonstrated that in regions far away from the ERC, noise was effectively removed however, at the cost of detail within the prostate. ACER and ANLM were more effective in retaining the prostatic detail, with ACER having an average improvement over the uncorrected slice of $11.7$ dB. Similar to the background SNR results, ACER had the highest average CNR with ROVST in second. ACER demonstrated an average $11.2$ dB improvement over the uncorrected slice in CNR. Subsequent p-value analysis (Table~\ref{tab:pval_real}) showed the average improvement over the uncorrected slices for each approach was statistically significant with p-values of less than $0.05$.

\begin{table*}[htbp]
  \centering
  \caption{The patient experiment CNR of two regions and the SNR of a background region are shown (largest values are shown in bold). ACER demonstrates an average increase of $11.7$ dB and $11.2$ dB for SNR and CNR respectively over the uncorrected (UC) slice which has no noise suppression applied.}
      \begin{tabular}{c|ccccc|ccccc}
    \toprule
    \textbf{} & \multicolumn{5}{c}{\textbf{Background SNR}} & \multicolumn{5}{c}{\textbf{CNR}} \\
    \midrule
    \textbf{Case} & \textbf{ACER} & \textbf{ROVST} & \textbf{LMMSE} & \textbf{ANLM} & \textbf{UC} & \textbf{ACER} & \textbf{ROVST} & \textbf{LMMSE} & \textbf{ANLM} & \textbf{UC} \\
    \midrule
    1     & \textbf{34.2} & 22.6  & 31.5  & 23.1  & 19.4  & \textbf{37.0} & 25.3  & 34.5  & 25.8  & 22.1 \\
    2     & \textbf{26.8} & 21.7  & 25.7  & 21.1  & 18.5  & \textbf{31.3} & 26.2  & 30.7  & 25.7  & 23.0 \\
    3     & 33.1  & \textbf{34.7} & 32.2  & 26.6  & 19.6  & 36.1  & \textbf{37.7} & 35.3  & 29.6  & 22.6 \\
    4     & 32.3  & \textbf{36.3} & 34.8  & 27.7  & 20.2  & 30.6  & \textbf{34.6} & 33.3  & 26.0  & 18.4 \\
    5     & \textbf{34.1} & 33.4  & 32.5  & 26.7  & 22.1  & \textbf{29.1} & 28.4  & 27.6  & 21.7  & 17.2 \\
    6     & \textbf{34.8} & 31.6  & 33.1  & 26.5  & 19.9  & \textbf{30.9} & 27.9  & 29.5  & 22.7  & 16.1 \\
    7     & \textbf{34.1} & 33.2  & 33.3  & 25.7  & 22.3  & \textbf{33.6} & 32.7  & 32.9  & 25.2  & 21.8 \\
    8     & 32.6  & \textbf{36.2} & 33.9  & 27.3  & 19.5  & 35.6  & \textbf{39.2} & 37.0  & 30.3  & 22.5 \\
    9     & 33.6  & \textbf{35.0} & 34.5  & 27.4  & 19.6  & 34.6  & \textbf{36.0} & 35.7  & 28.5  & 20.7 \\
    10    & 34.0  & \textbf{37.0} & 35.7  & 27.6  & 19.6  & 33.0  & \textbf{36.2} & 35.1  & 26.7  & 18.8 \\
    11    & \textbf{33.0} & 27.9  & 25.4  & 21.0  & 13.8  & 36.1  & \textbf{41.0} & 39.2  & 34.0  & 26.8 \\
    12    & \textbf{26.5} & 23.5  & 20.3  & 20.1  & 13.2  & \textbf{26.8} & 23.8  & 20.7  & 20.4  & 13.5 \\
    13    & 28.1  & \textbf{28.2} & 19.0  & 22.3  & 13.3  & 24.8  & \textbf{24.9} & 15.9  & 19.0  & 10.0 \\
    14    & 22.8  & \textbf{24.9} & 20.4  & 21.4  & 13.7  & 22.3  & \textbf{24.5} & 20.3  & 21.1  & 13.4 \\
    15    & \textbf{25.9} & 23.9  & 18.4  & 20.7  & 13.2  & \textbf{26.6} & 24.6  & 19.4  & 21.4  & 13.8 \\
    16    & 25.4  & \textbf{25.6} & 21.5  & 21.5  & 14.1  & 24.7  & \textbf{24.9} & 21.2  & 20.9  & 13.5 \\
    17    & 24.8  & \textbf{25.7} & 19.6  & 21.4  & 13.4  & 24.0  & \textbf{25.0} & 19.3  & 20.7  & 12.6 \\
    18    & \textbf{19.2} & 13.1  & 16.8  & 15.0  & 12.8  & \textbf{17.9} & 11.7  & 15.5  & 13.6  & 11.4 \\
    19    & \textbf{18.7} & 13.3  & 16.0  & 15.5  & 12.6  & \textbf{18.3} & 12.8  & 15.7  & 15.1  & 12.2 \\
    20    & 12.9  & 11.9  & \textbf{13.8} & 13.0  & 11.7  & 10.6  & 9.5   & \textbf{11.5} & 10.6  & 9.3 \\
    \midrule
    \textbf{Avg.} & \textbf{28.3} & 27.0  & 25.9  & 22.6  & 16.6  & \textbf{28.2} & 27.3  & 26.5  & 22.9  & 17.0 \\
    \bottomrule
         \end{tabular}%
     \label{tab:snr_cnr_real}%
  \end{table*}%

\begin{table}[htbp]
  \centering
  \caption{The p-values for the metrics measured for the patient experiments. Values below $0.05$ indicate the average score for all slices corrected by each approach represents statistically significant change from the uncorrected slices. All approaches have p-values below $0.05$.}
    \begin{tabular}{r|rrrr}
    \toprule
    \textbf{Metric} & \multicolumn{1}{c}{\textbf{ACER}} & \multicolumn{1}{c}{\textbf{ROVST}} & \multicolumn{1}{c}{\textbf{LMMSE}} & \multicolumn{1}{c}{\textbf{ANLM}} \\
    \midrule
    \textbf{Background SNR} & 4.56E-11 & 1.30E-07 & 8.96E-09 & 8.93E-10 \\
    \textbf{CNR} & 1.54E-11 & 1.30E-07 & 5.99E-09 & 8.88E-10 \\
    \bottomrule
    \end{tabular}%
  \label{tab:pval_real}%
\end{table}%

The edge preservation (EP) measurement evaluates image edge degradation. The EP measurement compares the noise-free reconstruction with the uncorrected image and can be calculated as follows~\cite{Sattar}:

\begin{equation}\label{Eq9}
\Upsilon = \frac{\Sigma(\bigtriangledown^2V - \overline{\bigtriangledown^2V})\cdot(\bigtriangledown^2\hat G - \overline{\bigtriangledown^2\hat G})}{\sqrt{\Sigma(\bigtriangledown^2V - \overline{\bigtriangledown^2V})^2\cdot\Sigma(\bigtriangledown^2\hat G - \overline{\bigtriangledown^2 \hat G})^2}}
\end{equation}

\noindent where $\bigtriangledown^2{V}$ and $\bigtriangledown^2\hat G$ are the Laplacian of the intensity bias corrected image and noise-free reconstruction respectively using a $3\times3$ filter. The parameters, $\overline{\bigtriangledown^2V}$ and $\overline{\bigtriangledown^2\hat G}$, are the mean values of a neighbourhood around $\bigtriangledown^2V$ and $\bigtriangledown^2\hat G$. An image where there is perfect EP results in a measurement of $\Upsilon = 1$. This refers to the technique's ability to retain the structure and edges of the image. For the purpose of this study, since noise can be recognized as edges or details, the EP metric is calculated for the prostate gland only using a user defined mask. This region was selected for high SNR and high importance for detail preservation.
\begin{table}[htbp]
  \centering
  \caption{Patient experiment edge preservation results: ANLM has the highest average edge preservation (EP) metrics as a result of insufficient noise suppression. ROVST and LMMSE demonstrate lower average metrics as a result of overcompensation. ACER defines an optimal balance between noise suppression and edge preservation which enhances visualization with the second highest EP metrics.}
    \begin{tabular}{c|cccc}
    \toprule
    \textbf{Case} & \textbf{ACER} & \textbf{ROVST} & \textbf{LMMSE} & \textbf{ANLM} \\
    \midrule
    1     & 0.977 & 0.994 & 0.982 & \textbf{1.000} \\
    2     & 0.936 & 0.982 & 0.954 & \textbf{0.996} \\
    3     & 0.953 & 0.875 & 0.932 & \textbf{0.979} \\
    4     & \textbf{0.956} & 0.840 & 0.847 & 0.954 \\
    5     & 0.846 & 0.832 & 0.836 & \textbf{0.957} \\
    6     & 0.933 & 0.881 & 0.907 & \textbf{0.980} \\
    7     & 0.895 & 0.884 & 0.890 & \textbf{0.979} \\
    8     & 0.971 & 0.863 & 0.930 & \textbf{0.975} \\
    9     & 0.896 & 0.861 & 0.881 & \textbf{0.963} \\
    10    & 0.954 & 0.869 & 0.903 & \textbf{0.976} \\
    11    & 0.923 & 0.792 & 0.938 & \textbf{0.973} \\
    12    & 0.970 & 0.867 & 0.872 & \textbf{0.981} \\
    13    & 0.960 & 0.896 & 0.902 & \textbf{0.984} \\
    14    & 0.985 & 0.923 & 0.921 & \textbf{0.987} \\
    15    & 0.935 & 0.838 & 0.860 & \textbf{0.969} \\
    16    & 0.952 & 0.868 & 0.868 & \textbf{0.978} \\
    17    & 0.973 & 0.903 & 0.906 & \textbf{0.984} \\
    18    & 0.957 & 0.986 & 0.957 & \textbf{0.999} \\
    19    & 0.980 & 0.993 & 0.981 & \textbf{1.000} \\
    20    & 0.974 & 0.992 & 0.971 & \textbf{1.000} \\
    \midrule
    \textbf{Avg.} & 0.946 & 0.897 & 0.912 & \textbf{0.981} \\
    \bottomrule
    \end{tabular}%
  \label{tab:epm}%
\end{table}%

Considering the EP of the noise compensation approaches (Table~\ref{tab:epm}), ANLM had the highest average EP with ACER having the second highest. In the real T2 cases, noise was more prominent than compared to the phantoms and as a result, more compensation was required to suppress the noise. This led to overcompensation in other regions where detail is important. Following the conclusions made in the phantom experiment, ROVST and LMMSE led to over suppression of noise and a lower EP measurement. ANLM however, had better EP for all but one case, as a consequence of its insufficient noise compensation. Due to the strong presence of noise in these slices, the Laplacian operator of the EP metric realized noise as edges. The insufficient noise suppression by ANLM resulted in structure preservation in the prostate, however also retained noise in regions of low SNR. This was demonstrated by the lower average background SNR compared to ACER.  ACER proved to have a suitable balance of noise suppression and EP as a result of the non-stationary unified ERC parametric model used.

\begin{figure*}[ht!]
	\centering
	\includegraphics[width=1\linewidth]{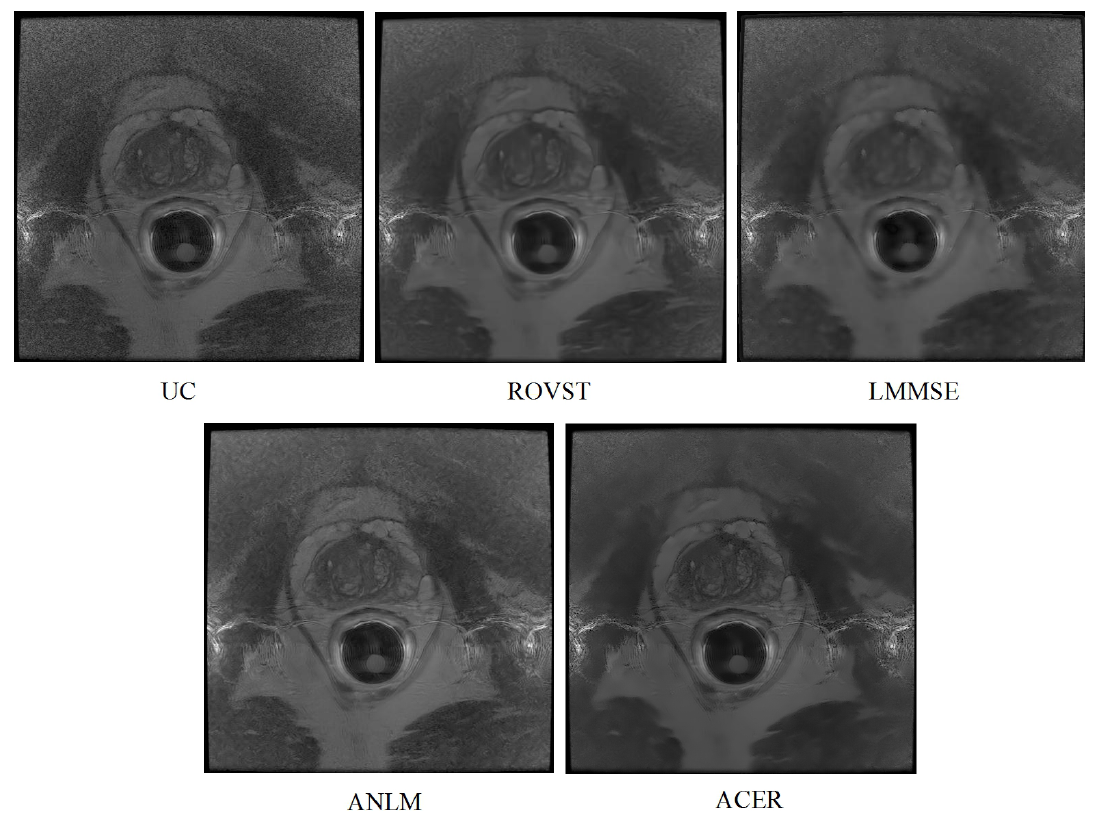}
	\caption{Case 12: A central T2 MRI slice from a patient imaged using a Hologic rigid ERC with moderate noise compensated by various approaches. ACER maintains the detail within the prostate while compensating for the background noise.}
	\label{fig:case12}
\end{figure*}

\begin{figure}[ht!]
	\centering
	\includegraphics[width=1\linewidth]{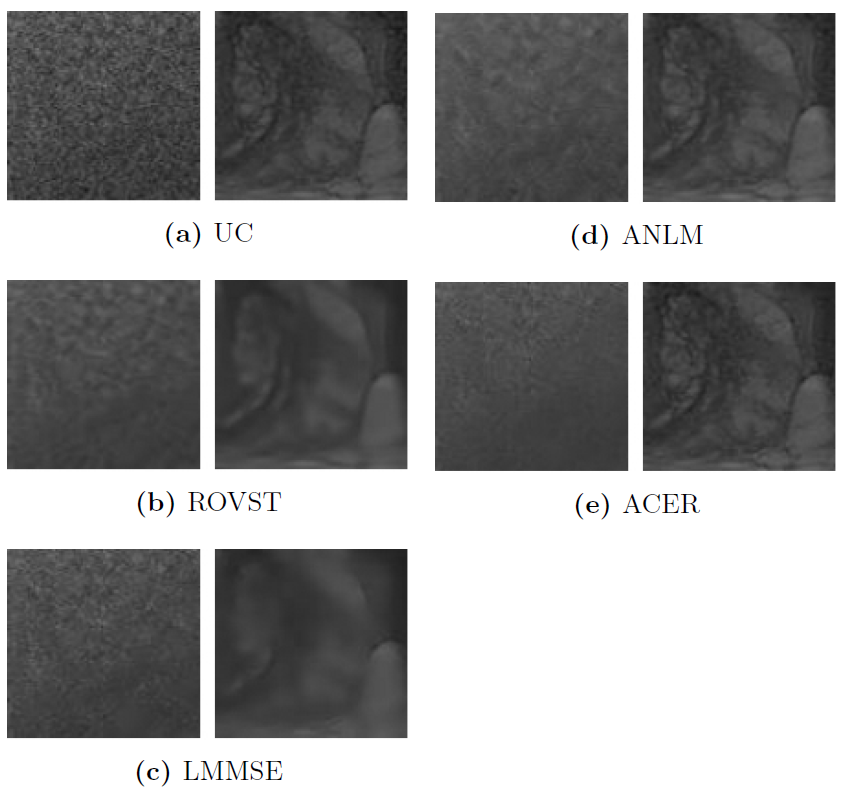}	
	\caption{Close-up views of background (left column) and prostate (right column) regions for Case 12. The selected regions are shown in Fig.~\ref{fig:case12_zoomregions}.}
	\label{fig:case12_zoom}
\end{figure}

\begin{figure*}[ht!]
	\centering
	\includegraphics[width=1\linewidth]{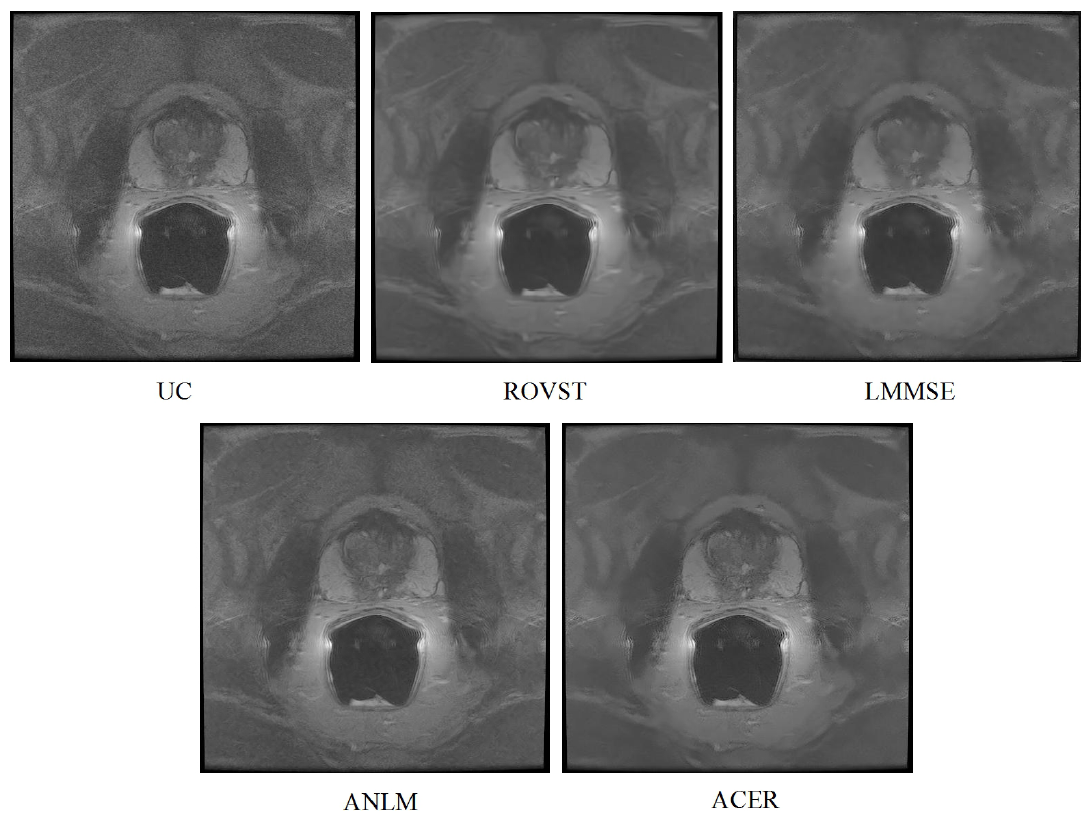}
	\caption{Case 3: A central T2 MRI slice from a patient imaged using a Medrad inflatable ERC compensated by various approaches. LMMSE and ROVST suppress noise in the background, consequently blurring details within the prostate. ACER effectively compensates the noise in low signal regions while taking advantage of the high signal near the coil. ANLM maintains similar detail preservation however retains some noise.}
	\label{fig:case3}
\end{figure*}

The EP analysis is further supported by the visual results shown in Fig.~\ref{fig:case12}, Fig.~\ref{fig:case12_zoom}, and Fig.~\ref{fig:case3}. LMMSE and ROVST are able to apply moderate noise suppression in the background regions where signal is low, however nearest the coil the prostate details are difficult to visualize due to overcompensation. ANLM is more effective in retaining the detail within the prostate region however at the cost of retaining the noise farther away from the ERC at high noise levels. ACER strikes an optimal balance between detail preservation within the prostate where signal is higher and effectively suppresses noise in the regions with low signal. This correction allows for improved visibility for diagnosis.

\begin{figure}[ht!]
	\centering
	\includegraphics[width=0.5\linewidth]{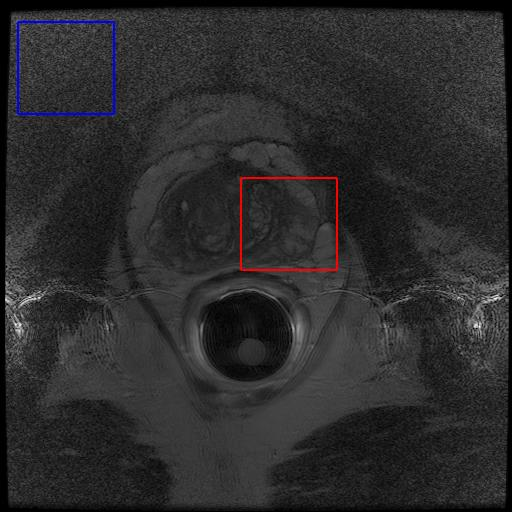}
	\caption{Selected background and prostate regions (shown on the uncorrected image) for closer inspection in Fig.~\ref{fig:case12_zoom}.}
	\label{fig:case12_zoomregions}
\end{figure}

\subsubsection{Image Analysis and Subjective Interpretation}
To appropriately assess the quality of the noise compensation approaches, a blind subjective scoring system similar to the evaluation system proposed by Walsh~et al.~\cite{Walsh2005} was used. In this system, the scorers were unaware of which approach was applied on the compensated data presented to them. A central slice from three volumes was selected and evaluated by seven evaluators ranging in experience. They are listed below from most to least experience:

\begin{itemize}
\item{MH, 16 years of clinical radiology experience with specialization in genitourinary cancers and 11 years of experience interpreting prostate MRI}
\item{LM, 7 years of clinical radiology experience with specialization in cancer imaging}
\item{FK, 5 years of prostate MRI research experience}
\item{HC, 1.5 years of clinical radiology experience}
\item{AM, 1.5 years of clinical imaging research experience}
\item{JK, 2 months of clinical prostate MRI experience}
\item{KC, 50 hours of clinical prostate MRI experience}
\end{itemize}

\begin{figure*}[htp]
	\centering
	\includegraphics[width=1\linewidth]{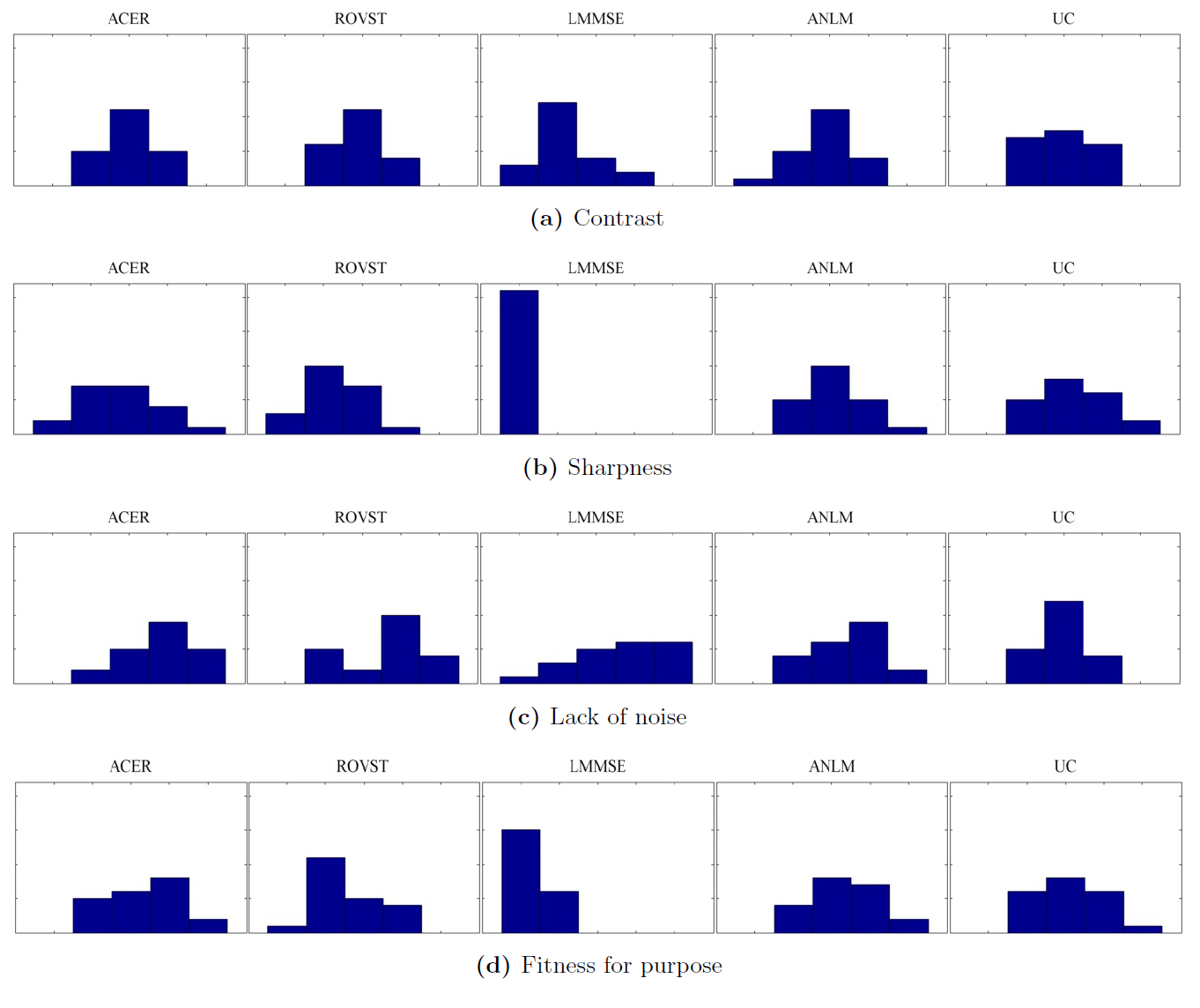}
	\caption[Subjective scoring histograms for the compared approaches]{Subjective scoring histograms for the compared approaches. The y-axis depicts Frequency (0 to 22) and the x-axis depicts the subjective score (1 to 5).}
	\label{fig:subscore}
\end{figure*}

To collect the subjective scores, the noise-suppressed and uncorrected versions of three slices were presented to the evaluators in an unknown and random sequence. Based on the individual slice, they were asked to assess the reconstruction based on the following criteria: contrast, sharpness, lack of noise and fitness for purpose. These criteria can be scored using the following terms: very poor, poor, satisfactory, good or very good. For the sake of our evaluation, we assigned these scores from $1$ to $5$, with $1$ being very poor and $5$ being very good. The rank sums~(Eq.~\ref{eq:ranksum}), median and F-pseudosigma scores~(Eq.~\ref{eq:fsigma}) across all slices and evaluators were calculated and are shown in Table~\ref{tab:subscore_ranksum}, Table~\ref{tab:subscore_median}, and Table~\ref{tab:subscore_fpseudosigma}. Histograms of each scoring criterion and the frequency of each score across all evaluators is included in Fig.~\ref{fig:subscore}.

\noindent The rank sum, $S_R$, is the total of all subjective scores by all the evaluators for a particular criterion.

\begin{equation}\label{eq:ranksum}
S_R = \sum_{i = 1}^{N}\sum_{j = 1}^{M} S_{ij},
\end{equation}

\noindent where $N$ is the number of evaluators and $M$ is the number of slices evaluators evaluated and $S_{ij}$ are the individual scores of each evaluator for each slice. The total rank sum can then be used to determine whether in general the evaluators decided a particular criterion was high or low for a given approach.

\noindent The next metric considered is the F-pseudosigma, $F_\sigma$, which is a measurement of variance and is calculated using:

\begin{equation}\label{eq:fsigma}
F_\sigma = \frac{IQR}{1.349},
\end{equation}

\noindent where IQR is the interquartile range. A smaller F-pseudosigma denotes a more precise score.

Considering the histograms~(Fig.~\ref{fig:subscore}), rank sum~(Table~\ref{tab:subscore_ranksum}), median~(Table~\ref{tab:subscore_median}) and F-pseudosigma~(Table~\ref{tab:subscore_fpseudosigma}) metrics for contrast, ACER had the highest rank sum with a median score of satisfactory. It also had the smallest F-pseudosigma which indicates there was little variation between all scores. For the sharpness criterion, it was interesting that the uncorrected image had the largest rank sum with ANLM having the next highest rank sum. ACER, ANLM and uncorrected tied with the highest median scores of satisfactory however also had the highest F-pseudosigmas indicating large variation in opinion. It was unanimous however that LMMSE had very poor sharpness and was found to be less sharp than the uncorrected slices. For the lack of noise criterion, ACER again had the largest rank sum with a median score of good. All correction approaches had high rank sums and median scores of good however again, F-pseudosigmas hinted at large variance in opinion. This may have been caused by the large number of evaluators and the variance in noise level between cases. Finally, ACER and ANLM had the highest rank sums for fitness for purpose with a median score of satisfactory. LMMSE and ROVST were found to be unfit for the purpose in comparison to uncorrected slices. It is intriguing to point out that evaluators found that the uncorrected slices were just as sufficient for analysis as ACER and ANLM however there was large variance in opinion with large F-pseudosigma scores.

\begin{table}[htbp]
  \centering
  \caption{The rank sum subjective score values (with highest scores shown in bold): ACER has the highest rank sum for contrast and lack of noise.}
    \begin{tabular}{c|ccccc}
    \toprule
    \textbf{Scoring Criterion} & \textbf{ACER} & \textbf{ROVST} & \textbf{LMMSE} & \textbf{ANLM} & \textbf{UC} \\
    \midrule
    \textbf{Contrast} & \textbf{63} & 61    & 47    & 60    & 62 \\
    \textbf{Sharpness} & 58    & 48    & 21    & 65    & \textbf{68} \\
    \textbf{Lack of noise} & \textbf{80} & 76    & 76    & 72    & 62 \\
    \textbf{Fitness for purpose} & \textbf{70} & 54    & 27    & \textbf{70} & 65 \\
    \bottomrule
    \end{tabular}%
  \label{tab:subscore_ranksum}%
\end{table}%

\begin{table}[htbp]
  \centering
  \caption{The median subjective score values (with the highest scores shown in bold): ACER and ANLM demonstrated the same median scores as UC except for lack of noise where all approaches improved upon UC.}
    \begin{tabular}{c|ccccc}
    \toprule
    \textbf{Scoring Criterion} & \textbf{ACER} & \textbf{ROVST} & \textbf{LMMSE} & \textbf{ANLM} & \textbf{UC} \\
    \midrule
    \textbf{Contrast} & \textbf{3} & \textbf{3} & 2     & \textbf{3} & \textbf{3} \\
    \textbf{Sharpness} & \textbf{3} & 2     & 1     & \textbf{3} & \textbf{3} \\
    \textbf{Lack of noise} & \textbf{4} & \textbf{4} & \textbf{4} & \textbf{4} & 3 \\
    \textbf{Fitness for purpose} & \textbf{3} & 2     & 1     & \textbf{3} & \textbf{3} \\
    \bottomrule
    \end{tabular}%
  \label{tab:subscore_median}%
\end{table}%

\begin{table}[htbp]
  \centering
  \caption{The F-pseudosigma subjective score values (with the lowest scores shown in bold): With the exception of the unanimous decision that LMMSE had poor sharpness, most of the criteria for the approaches had high variance indicating large inconsistencies in opinion implying that personal preference has a large impact upon the approach.}
    \begin{tabular}{c|ccccc}
    \toprule
    \textbf{Scoring Criterion} & \textbf{ACER} & \textbf{ROVST} & \textbf{LMMSE} & \textbf{ANLM} & \textbf{UC} \\
    \midrule
    \textbf{Contrast} & \textbf{0.37} & 0.74  & 0.74  & 0.74  & 1.48 \\
    \textbf{Sharpness} & 0.93  & 0.74  & \textbf{0.00} & 0.93  & 0.93 \\
    \textbf{Lack of noise} & 0.93  & 0.93  & 1.48  & 0.74  & \textbf{0.19} \\
    \textbf{Fitness for purpose} & 0.93  & \textbf{0.74} & \textbf{0.74} & \textbf{0.74} & 1.48 \\
    \bottomrule
    \end{tabular}%
  \label{tab:subscore_fpseudosigma}%
\end{table}%

\subsubsection{Visual Analysis}
Visual results for two different cases are shown in Fig.~\ref{fig:case12} and Fig.~\ref{fig:case3} for a Hologic rigid ERC and a Medrad inflatable ERC respectively. The results demonstrate ACER's ability to retain prostate detail with effective compensation of background noise using different ERCs with different SNR characteristics. In Fig.~\ref{fig:case12}, it is evident that LMMSE and ANLM are able to reduce the noise in the background regions however with noise still visibly present. ROVST does a better job at compensating for noise however upon closer inspection of Fig.~\ref{fig:case12} in regions specified by Fig.~\ref{fig:case12_zoom} for a background and prostate region (Fig.~\ref{fig:case12_zoomregions}) it is apparent that the level of detail is compromised for these approaches. ROVST and LMMSE approaches were unable to preserve the tissue texture within the prostate, demonstrating oversmoothing in the prostate in order to compensate for the high level of noise in the background. In contrast, ANLM was able to retain the detail within the prostate however showed some noise in the background. ACER successfully balances the noise reduction and the detail preservation by incorporating the ERC SNR profile as well as the non-stationary characteristics of the MRI data. Similar conclusions can be made when considering the performance of the approaches for the inflatable ERC case (Fig.~\ref{fig:case3}). In this example, the ANLM applies insufficient noise compensation and shows evidence of noise. ROVST and LMMSE suppress the noise however at the cost of removing detail in the prostate. Again, ACER exhibits apt noise compensation while retaining tissue texture and details.

\subsection{Timing Analysis}
The various MRI compensation approaches were also analyzed based on their computation times. Tests were completed on a $3.10$ GHz AMD Athlon(tm) II X3 445 processor with $4.00$ GB of RAM. The various approaches were not optimized for timing performance. The timing analysis is shown for the patient data in Table~\ref{tab:time}. The LMMSE approach demonstrated the fastest computation times with an average computation time of $0.13$ s while ANLM exhibited the slowest computation time with an average calculation time of  $1060$ s. The proposed approach, ACER, showed middle range performance with an average computation time of $284$ s.

\begin{table}[htbp]
  \centering
  \caption{Computation times for each approach on the real T2 endorectal MRI shown in seconds. Shortest computation times are shown bolded. LMMSE had the shortest average computation time with $0.13$ s while ANLM had the longest average computation time with $1060$ s.}
    \begin{tabular}{c|cccc}
    \toprule
    \textbf{Case} & \textbf{ACER} & \textbf{ROVST} & \textbf{LMMSE} & \textbf{ANLM} \\
    \midrule
    1     & 370   & 9.82  & \textbf{0.17} & 1170 \\
    2     & 265   & 8.22  & \textbf{0.12} & 1090 \\
    3     & 256   & 8.33  & \textbf{0.13} & 1310 \\
    4     & 270   & 8.43  & \textbf{0.13} & 1340 \\
    5     & 268   & 8.47  & \textbf{0.11} & 1310 \\
    6     & 274   & 8.34  & \textbf{0.12} & 1280 \\
    7     & 299   & 8.32  & \textbf{0.12} & 1330 \\
    8     & 361   & 8.43  & \textbf{0.13} & 1330 \\
    9     & 295   & 8.5   & \textbf{0.12} & 1240 \\
    10    & 285   & 8.42  & \textbf{0.12} & 1210 \\
    11    & 272   & 8.31  & \textbf{0.12} & 1210 \\
    12    & 276   & 7.22  & \textbf{0.19} & 923 \\
    13    & 275   & 6.99  & \textbf{0.11} & 888 \\
    14    & 274   & 6.85  & \textbf{0.12} & 848 \\
    15    & 273   & 7.03  & \textbf{0.12} & 846 \\
    16    & 273   & 6.99  & \textbf{0.12} & 789 \\
    17    & 272   & 7.06  & \textbf{0.14} & 871 \\
    18    & 272   & 7.28  & \textbf{0.13} & 753 \\
    19    & 272   & 7.34  & \textbf{0.14} & 693 \\
    20    & 273   & 7.28  & \textbf{0.12} & 690 \\
    \midrule
    \textbf{Avg.} & 284   & 7.88  & \textbf{0.13} & 1060 \\
    \bottomrule
    \end{tabular}%
  \label{tab:time}%
\end{table}%

\section{Conclusion}
\label{Conclusions}
In this study, a novel noise compensation approach for coil intensity corrected endorectal MRI images is presented. Adaptive Coil Enhancement Reconstruction (ACER) uses a spatially-adaptive Monte Carlo sampling approach to estimate the Rician-distributed posterior in MRI images to reconstruct the noise compensated image. ACER takes advantage of the known SNR characteristics of an ERC to develop a non-spatial unified ERC parametric model that models the SNR profile presented by the ERC. This allows for effective noise suppression and detail preservation in the prostate.  This approach to noise compensation for coil intensity corrected endorectal MRI images is particularly useful for retrospective studies where the original raw data is not available and only the coil intensity corrected data is accessible.  Experimental results using both phantom and patient data showed that ACER provided strong performance in terms of SNR, CNR and edge preservation when compared to a number of existing approaches.  Future work involves extending ACER to automatically estimate the SNR profile of the ERC, thus eliminating the necessity for the ERC SNR profile, investigating the effect and efficacy of ACER on improving the quality of multi-sequence endorectal modalities such as correlated diffusion imaging~\cite{CDI,DCDI}, investigating the extension of ACER for endorectal compressed sensing MRI~\cite{sparse}, and investigating its efficacy for improving cancer detection~\cite{Farzad1,Farzad2,Farzad3}.

\section*{Acknowledgements}
This research was undertaken, in part, thanks to funding from the Canada Research Chairs program. The study was also funded by the Natural Sciences and Engineering Research Council (NSERC) of Canada and the Ontario Ministry of Research and Innovation. We would also like to thank our subjective scorers for helping us complete subjective scoring.

\section*{Competing Interests}
The authors have declared that no competing interests exist.

\section*{Author Contributions}
Conceived and designed the methodology: DL, AW. Performed the experiments: DL, AW, AM. Analyzed the data: DL, AW, MH. Wrote the paper: DL, AW, AM, MH.

\bibliography{mridenoise}



\end{document}